\documentclass[aps,pra,amsfonts,amsmath,amssymb,preprint]{revtex4-1}
\usepackage[utf8]{inputenc}
\usepackage{textcomp}
\usepackage{appendix}
\usepackage{graphicx}

\begin{document}

\title{Dynamics within a tunable harmonic/quartic waveguide}
\author{Rudolph N. Kohn, Jr.}
\email{rudy.kohn@sdl.usu.edu}
\author{James A. Stickney}
\affiliation{Space Dynamics Laboratory, Albuquerque, New Mexico 87106, USA}

\begin{abstract}

We present an analytical solution to the dynamics of a noninteracting cloud of 
thermal atoms in a cigar-shaped harmonic trap with a quartic perturbation along 
the axial direction.  We calculate the first and second moments of position, 
which are sufficient to characterize the trap. The dynamics of the thermal cloud 
differ notably from those of a single particle, with an offset to the 
oscillation frequency that persists even as the oscillation amplitude approaches 
zero.  We also present some numerical results that describe the effects of 
time-of-flight on the behavior of the cloud in order to better understand the 
results of a hypothetical experimental realization of this system.

\end{abstract}

\maketitle

\section{Introduction} \label{sec:intro}

Cold atom interferometers have proven themselves to be valuable tools for 
examining a variety of effects, leveraging the wave nature of matter and the 
large rest mass of atoms to make extremely precise measurements.  They have been 
used to measure gravity \cite{Fixler2007, Peters2001, Bertoldi2006}, multi-axis 
accelerations \cite{Canuel2006, Wu2017, Dickerson2013}, rotations 
\cite{Burke2009, Wu2007, Dutta2016}, electric polarizability \cite{Deissler2008}, 
fundamental quantities such as the fine structure constant \cite{Bouch2011}, and 
to test predictions of general relativity \cite{Chung2009}.  In general, these 
interferometers use controlled light pulses to apply coherent momentum kicks to separate clouds into subsets with different momenta and reflect them back 
toward each other.  Between kicks, the atoms evolve in free space or along 
a waveguide.  In most of the guided atom interferometers, laser pulses reflect 
the atoms long before the confinement turns the atoms around \cite{Wu2007, 
Deissler2008, Burke2009}, but there is at least one example of an interferometer 
allowing the atoms to complete a full oscillation in the confinement 
potential \cite{Horikoshi07}.

Interferometers produced from trapped atoms can be smaller than their free space 
counterparts, and the trapping potentials can hold the atoms against gravity, 
but the existence of a preferred axis of oscillation makes precise alignment of 
the excitation beams critical \cite{Burke2009}.  Using the potential to reflect 
the clouds means that, for a given separation, the interrogation time can be 
longer. However, uncontrolled variation in the trap potential can cause unwanted 
effects, and interactions between atoms can also weaken the signal
\cite{Horikoshi07}. In general, trapping potentials will always depart somewhat 
from perfect harmonicity, if only because of errors introduced in the 
fabrication process or inherent in the trap design.  Therefore, it is extremely 
valuable to be able to identify and compensate for unwanted deviations in the 
trap shape, especially if it can be done without altering the hardware.

To this end, we developed a method for producing atom chip traps which permit 
fine control of the potential along the axis of a magnetic waveguide.  Several 
polynomial terms can be controlled by tuning the currents through 
several pairs of wires and the spacing of the wires can minimize higher order 
contributions to the potential \cite{Stickney14}. These tunable atom chip 
waveguides allow us, in theory, to produce carefully tailored potentials, but 
imperfections in the manufacturing process call for fine-tuning in order to 
approach the desired potential as closely as possible.  Therefore, it is crucial 
to examine the dynamics of atoms in the potentials and have a clear theoretical 
understanding of the effects of deviations in the potential.

In this paper, we will examine the dynamics of a cold cloud of atoms in a 
one-dimensional trap with harmonic and quartic components.  We will assume the 
other two axes are well-confined, harmonic, conservative, and separable.  The 
harmonic and quartic terms are particularly interesting because the harmonic 
term is solvable and the quartic term is often the leading unwanted contribution 
in trapped atom interferometers \cite{Horikoshi07}.  With a few minor 
approximations, we will derive solutions for the behavior of single particles 
and ensembles and show that there are qualitative differences between them.

The purely harmonic case can be thoroughly described using Boltzmann's kinetic 
theory. One of the more counterintuitive results involves the behavior of the 
clouds at long times.  While it might be assumed that a system of atoms in a 
perfectly harmonic trap would eventually reach some kind of thermal equilibrium, 
certain excitation modes, such as the monopole breathing mode, actually persist 
indefinitely, even in the presence of isotropic, energy-independent, elastic 
collisions \cite{GO99}.  The persistence of the monopole breathing mode was 
demonstrated experimentally in a system of cold atoms by \citet{Lobser15} in 
2015. Some simple substitutions show that, in addition to monopole breathing, 
center-of-mass oscillations of a small cloud along the weak axis of a 
cylindrically symmetric harmonic trap will also persist indefinitely, so long as 
the axes are separable, and collisions are isotropic, elastic, and 
energy-independent \cite{GO99}.  We will refer to such oscillations as 
``sloshing'' henceforth.  As we will show below, the addition of a quartic 
perturbation has several effects.  Most importantly, the quartic perturbation 
causes initially close atoms with slightly different energies to gradually 
separate, effectively randomizing their phases and resulting in the gradual 
decay of sloshing and the spreading out of the cloud in a quasi-thermal state at 
the center of the trap, even in the absence of collisions.  We note that this 
quasi-thermalization, or ``dephasing'' can be used to characterize the 
anharmonic contributions to the trapping potential. In one of our tunable atom 
chips, the parameters can be adjusted to minimize dephasing and iteratively 
approach a perfectly harmonic potential.  In addition, we will see below that 
the quartic contribution alters the frequency of the trap for clouds of atoms, 
even at infinitesimally small sloshing amplitudes.  Finally, the use of two 
independent parameters to describe our traps necessitates the measurement of two 
independent characteristics of the cloud.   The center of mass position, 
$\langle x \rangle$ and the size of the cloud $\sigma$, requiring two moments of 
position, are sufficient to uniquely determine the shape of such a trap.

The paper is divided into several sections.  In Section 
\ref{sec:theory}, we will proceed through the analytical solution of the
one-dimensional harmonic-quartic trap.  Section \ref{sec:results} describes
the results of the theory and examines some of the finer details.  In Section
\ref{sec:num}, we will use numerical methods to calculate the behavior of the 
clouds after some time of flight, which is a common technique used to observe
clouds of cold atoms. Finally, we will summarize our conclusions and describe
some future paths for inquiry in Section \ref{sec:conc}.

\section{Theory} \label{sec:theory}

In this paper, we will assume that the motion of the atomic gas in the $x$ 
direction can be decoupled from the pure harmonic motion in the $y$ and $z$ directions, i.e. 
$H_t = H(x,p_x) + H_\perp(y,z,p_y,p_z)$.  The trap along the x-axis is mostly 
harmonic, and the most significant deviation is a quartic term.  Thus, the 
Hamiltonian 
\begin{equation}
    H = \frac{p^2}{2m} + m \omega^2 \left( 
    \frac{x^2}{2} + \frac{x^4}{4 x_4^2} 
    \right)
 \label{eqn:quartham}
\end{equation}
governs the dynamics, with $m$ as the atomic mass, $\omega$ as the harmonic trap 
frequency, and $x_4^2$ describing the quartic contribution. 
We have also used $p$ as shorthand for $p_x$ and will continue to do so going 
forward.  $x_4^2$ can be 
either positive or negative, and its magnitude corresponds to the value of $x$
where the forces due to the harmonic and quartic terms are equal. 

In the limit where the oscillation amplitude $A$ is small, i.e. $A \ll |x_4|$, 
the dynamics of a particle in the Hamiltonian given by Eq. (\ref{eqn:quartham}) 
can be approximated by a sinusoid with an amplitude-dependent 
frequency.  The position of the particle is approximately
\begin{equation}
    \frac{x(t)}{A} = 
      \cos(\Omega t - \phi_0) + \frac{1}{32} \frac{A^2}{x_4^2}
 \cos(3 \Omega t - 3 \phi_0)
 \label{eqn:1particle}
\end{equation}
where $A$ is the amplitude and $\phi_0$ is the initial phase of the particle.  
The amplitude-dependent frequency is
\begin{equation}
    \frac{\Omega}{\omega} = 1 + \frac{3}{8} \frac{A^2}{x_4^2},
 \label{eqn:Omega}
\end{equation}
and does not depend on the initial phase. We neglect the part of the dynamics 
which oscillates at the frequency $3 \Omega$ in later calculations because
its magnitude is extremely small compared to the $\Omega$ term.

The effect of the perturbation on the dynamics of a single particle becomes 
dependent on only the amplitude of oscillation.  It is convenient to recast this 
in terms of the unperturbed energy
\begin{equation}
    E = \frac{p_0^2}{2 m} + \frac{1}{2} m \omega^2 x_0^2,
    \label{energy}
\end{equation}
where $x_0$ and $p_0$ are the initial coordinate and momentum.

\begin{figure}
 \includegraphics[width=17.2cm]{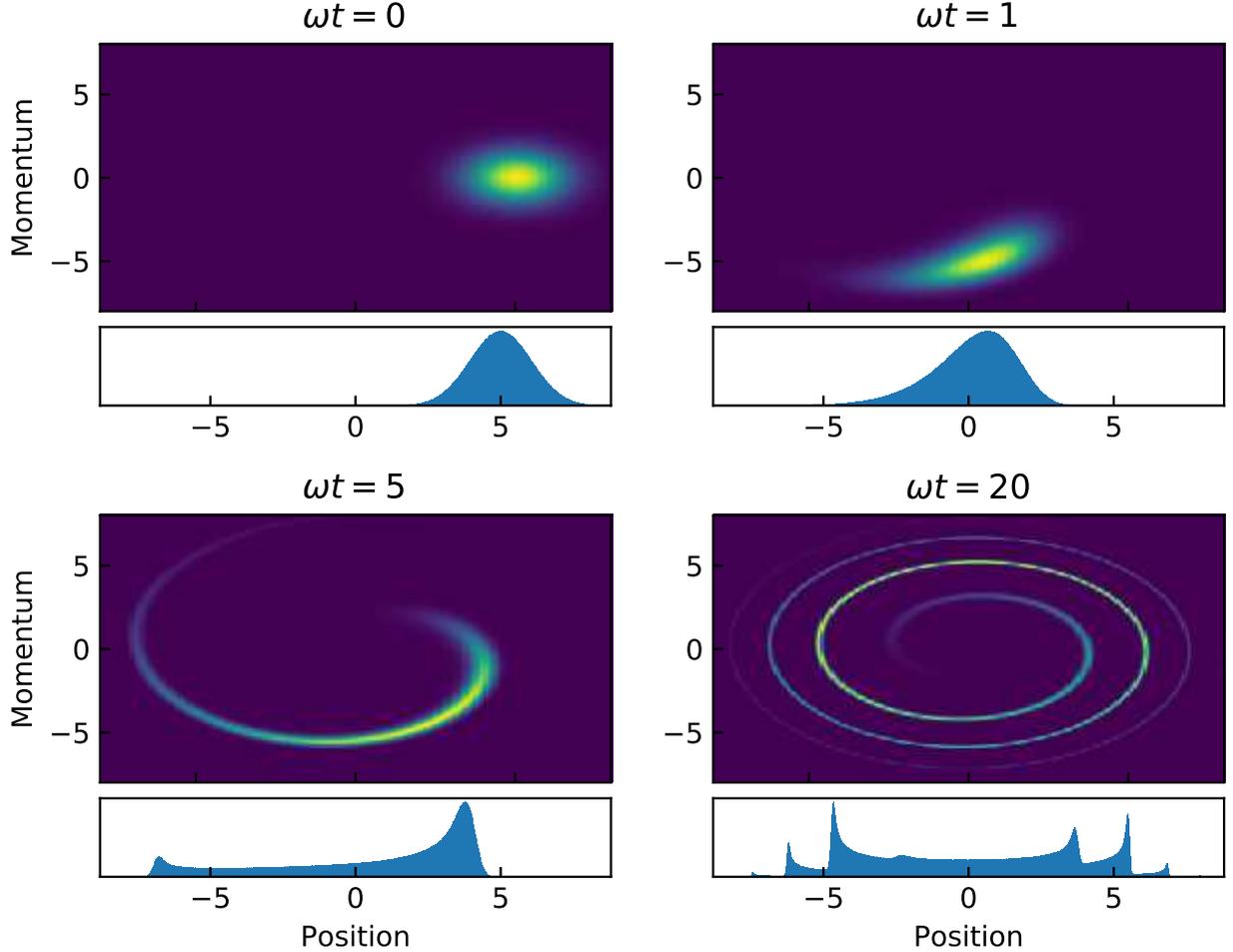}
 \caption{Diagrams of an atomic cloud as dephasing causes quasi-thermal 
equilibrium.  The figure is separated into four parts, with the top subplot 
showing the phase space density of the cloud at a given time, and the bottom 
subplot shows the density distribution in space only.  As time passes, atoms 
with small differences in energy lose phase coherence until the cloud eventually 
spreads out along the bottom of the trap, almost as if it has thermalized,
even though the model contains no interactions at all.} 
\label{fig:dephase}
\end{figure}

For the remainder of the paper, we will describe the dynamics of a single 
particle to be 
\begin{equation}
    x(t) = \sqrt{\frac{2E}{m \omega^2}} \cos(\Omega t - \phi_0),
    \label{eq:1part}
\end{equation}
where $\tan \phi_0 = p_0/m \omega x_0$ and 
\begin{equation}
    \frac{\Omega}{\omega} = 1 + \frac{3}{8}\frac{2 E}{m \omega^2 x_4^2}.
    \label{eq:Omega}
\end{equation}
From these expressions, we see that any particle can be placed on a constant 
energy ellipse in phase space and traces out that ellipse with a frequency that 
depends only on energy, independent of the initial phase of the oscillation.  
Fig.~\ref{fig:dephase} illustrates this effect by plotting the phase space
distribution of a cloud in a harmonic-quartic trap at several different times.

A classical cloud of atoms can be described by its phase space density, $f(x, p, 
t)$, where the phase space density describes the probability density to find an 
atom with the position $x$ and momentum $p$ at time $t$. The  n$^{\mathrm{th}}$ 
moment of position of a cloud of atoms is
\begin{equation}
 \langle x^n \rangle = \int_{-\infty}^{\infty} 
 \int_{-\infty}^{\infty} dx dp \, x^n f(x,p,t).
\end{equation}
In general, the phase space density $f(x, p, t)$ evolves in time, but in a 
non-interacting system each element of phase space moves independently from all 
others and can be tracked as it evolves.  By backtracking each point and its 
phase space density back to $t=0$, one can remove the dynamics from $f(x,p,t)$ 
and incorporate them into $x^n$ such that
\begin{equation}
 \langle x^n \rangle = \int_{-\infty}^{\infty} 
 \int_{-\infty}^{\infty} dx dp \, x_R^n(x,p,t) f_0(x,p),
    \label{moments}
\end{equation}
where $x_R(x,p,t)$ is the backtracked position at $t=0$ that the 
position at $(x,p,t)$ now represents.

To simplify the integration, it is convenient to transform Eq.(\ref{moments}) 
into a polar coordinate system such that 
\begin{equation}
x = \sqrt{2 \xi} \sigma_x \cos \phi \: \mathrm{and} \: 
p = \sqrt{2 \xi} \sigma_p \sin \phi,
\end{equation}
where $\xi = E / k_B T$ is the ratio between 
phase space energy, Eq. (\ref{energy}), and thermal energy, where $T$ is the 
initial temperature of the cloud.  Both coordinate and momentum are scaled by 
the thermal standard deviations, 
\begin{equation}
\sigma_p = \sqrt{m k_B T} \: \mathrm{and} \: \sigma_x =  \sqrt{k_B T / m \omega^2},  
\end{equation}
to make them unitless. The angle $\tan(\phi) = \sigma_x p/\sigma_p x$ is the initial polar angle of some point $(x,p)$ at $t=0$. In this new coordinate system
\begin{equation}
x_R = \sqrt{2 \xi} \sigma_x \cos(\Omega t + \phi),
\end{equation}
and Eq. (\ref{moments}) 
becomes
\begin{equation}
    \frac{\langle x^n \rangle }{(\sqrt{2} \sigma_x)^n} = \int_{0}^{\infty} 
    \int_{0}^{2 \pi} d\xi d\phi \sigma_x \sigma_p \xi^{n/2} 
    \cos^n(\Omega t + \phi) f_0(\xi, \phi),
    \label{mom}
\end{equation}
where $f_0(\xi,\phi)$ is simply the initial phase space density distribution 
converted to the new coordinate basis.

First consider the case where the atomic cloud is prepared in thermodynamic 
equilibrium at temperature $T$, in the trapping potential with frequency 
$\omega$.  At time $t=0$ the cloud is given a kick, resulting in initial
center of mass conditions
\begin{equation} 
p_D = \sqrt{2 \xi_D} \sigma_p \sin \phi_D \: \mathrm{and} \: 
x_D = \sqrt{2 \xi_D} \sigma_x \cos \phi_D. 
\end{equation}

$\xi_D$ and $\phi_D$ are defined with respect to $x_D$ and $p_D$ 
in the same way that $\xi$ and $\phi$ are defined in terms of $x$ and $p$, such that 
\begin{equation}
 \xi_D = \frac{x_D^2}{2 \sigma_x^2} + \frac{p_D^2}{2 \sigma_p^2} \textrm{ and } 
 \tan(\phi_D) = \sigma_x p_D/\sigma_p x_D.
\end{equation}
Starting from a thermal distribution in the harmonic trap, we displace it by
$x_D$ and $p_D$, and convert to the polar coordinate system, leading to
\begin{equation}
    f_0 = \frac{1}{2 \pi \sigma_x \sigma_p} \exp\left( -\xi - \xi_D + 2 
    \sqrt{\xi \xi_D} \cos(\phi - \phi_D) \right).
    \label{dis}
\end{equation}
Substituting Eq. (\ref{dis}) into Eq. (\ref{mom}) and converting the integral
to the new variables yields
\begin{equation}
    \frac{\langle x^n \rangle}{(\sqrt{2} \sigma_x)^n} = \frac{e^{-\xi_D}}{2 \pi} \int_{0}^{\infty} d\xi 
    \xi^{n/2}
    e^{-\xi} \int_{0}^{2 \pi} d\phi
    \cos^n(\Omega t + \phi_D + \phi) 
    \exp\left(2 \sqrt{\xi \xi_D} \cos \phi  \right).
    \label{eq:moments}
\end{equation}
The trigonometric power law reduction formula, $\cos^n \theta = \sum_m c_{nm} 
\cos m \theta$,  permits further simplification. Analytic expressions for the 
coefficients $c_{nm}$ can be found in \citet{GR07}, on page 31.  In this case, 
the three elements of interest are $c_{11} = 1$, $c_{20} = 1/2$, and 
$c_{22} = 1/2$.

Perhaps surprisingly, Equation~(\ref{eq:moments}) can be solved analytically 
without further approximation.  To simplify 
the notation, the moments can be rewritten as
\begin{equation}
    \frac{\langle x^n \rangle}{(\sqrt{2} \sigma_x)^n} = 
    \frac{1}{2}
    \sum_m c_{nm} \left( {\Upsilon}_{nm} + c.c  \right),
\end{equation}
with the $\Upsilon_{nm}$ given as
\begin{equation}
    {\Upsilon}_{nm} 
    =
    e^{-\xi_D + i m (\omega t + \phi_D)}
    \int_{0}^{\infty} d\xi \xi^{n/2}
    \exp\left[ - \left( 1 - i m \Lambda t \right) \xi \right]
    I_m(2 \sqrt{\xi \xi_D}),
    \label{integral}
\end{equation}
where $\Lambda \equiv 3 \omega \sigma_x^2 / 8 x_4^2$ and $I_m$ is the modified 
Bessel function of the first kind. The integrals in $\Upsilon_{20}$, 
$\Upsilon_{11}$ and $\Upsilon_{22}$ must be solved to calculate closed-form 
expressions for $\langle x \rangle$ and $\langle x^2 \rangle$. The integrals in 
$\Upsilon_{11}$ and $\Upsilon_{22}$ can be solved in terms of \citet{GR07}, 
section 6.631, equation 4. $\Upsilon_{20}$ can be solved in terms of equation 1 
in the same section. The results are
\begin{equation}
    \Upsilon_{20} = 1 + \xi_D,
    \label{eq:ups20}
\end{equation}
\begin{equation}
    \Upsilon_{11} = \frac{\sqrt{\xi_D}}{(1-i \Lambda t)^2}
    \exp \left[ -\xi_D + \frac{\xi_D}{(1-i \Lambda t)} + 
    i(\omega t + \phi_D) \right],
    \label{eq:ups11}
\end{equation}
and
\begin{equation}
    \Upsilon_{22} = \frac{\xi_D}{(1-2 i \Lambda t)^3}
    \exp \left[ -\xi_D + \frac{\xi_D}{(1-2 i \Lambda t)} +
    i (2 \omega t + 2 \phi_D) \right].
    \label{eq:ups22}
\end{equation}

With these solutions, it is possible to calculate the center-of-mass position 
and size of the cloud as a function of time for any chosen set of parameters 
that satisfies the requirement that the initial kick produce a maximum 
displacement in position much less than $|x_4|$.

\section{Results} \label{sec:results}

Having calculated the values of $\Upsilon_{nm}$, we can combine them into
expressions for physical properties that can be easily measured in an
experimental realization of this system.  The center-of-mass position and
size of the cloud are, in general, easily observed and calculated from
absorption images.

\begin{figure}
 \includegraphics[width=8.6cm]{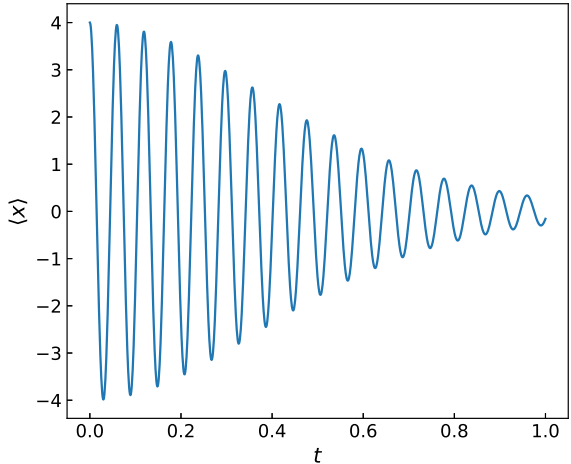}
 \caption{$\langle x \rangle$ as a function of time.  Most of the decay
 to quasi-thermal equilibrium is shown.  The plot uses parameters
 $\xi_D = 8$, $\phi_D = 0$, $\omega = 100$, $\sigma_x(t = 0) = 1$, and 
 $x_4 = 8$.  For these parameters, $\Lambda \cong 0.586$.}
 \label{fig:x}
\end{figure}

\begin{figure}
 \includegraphics[width=8.6cm]{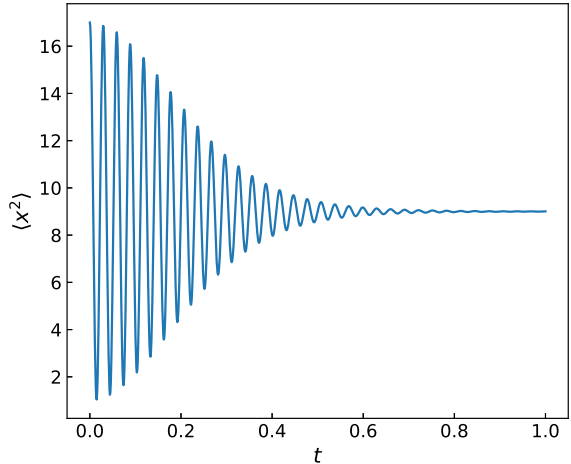}
 \caption{$\langle x^2 \rangle$ as a function of time for the same parameters
 as used in Figure~\ref{fig:x}.  Note the more rapid oscillations as well as
 the more rapid decay of the oscillations.}
 \label{fig:x2}
\end{figure}

\begin{figure}
 \includegraphics[width=8.6cm]{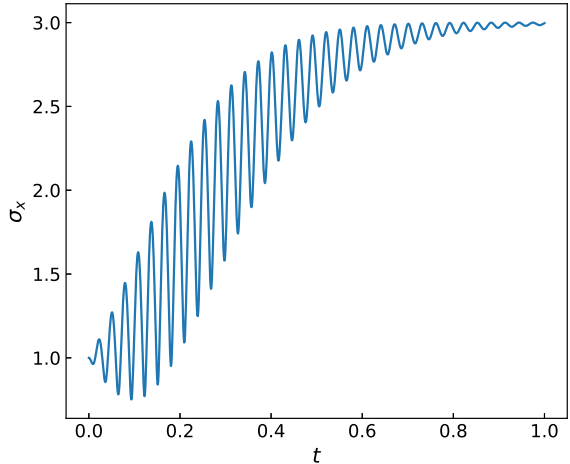}
 \caption{The size of the cloud as a function of time, using the same
 parameters as Figures~\ref{fig:x} and \ref{fig:x2}.}
 \label{fig:sig}
\end{figure}

$\Upsilon_{11}$ leads to the center-of-mass position
of the atomic cloud as a function of time.
\begin{equation}
 \langle x \rangle = \frac{\sigma_x \sqrt{2 \xi_D}}{[1+(\Lambda t)^2]^2}
 \exp \left[ \frac{-(\Lambda t)^2 \xi_D}{1 + (\Lambda t)^2} \right]
 \left( \cos(\Phi_1) 
 - 2 \Lambda t \sin(\Phi_1)
 - (\Lambda t)^2 \cos(\Phi_1) \right),
 \label{eq:mom1}
\end{equation}
where $\Phi_1 = \omega t + \phi_D + \Lambda t \xi_D / (1 + (\Lambda t)^2)$.  The first moment
resembles a decaying sinusoid.  The initial decay is Gaussian, but
at later times the denominator of the first term takes over.
$\Upsilon_{20}$ and $\Upsilon_{22}$ combine to give the second moment
\begin{equation}
\begin{aligned}
 \langle x^2 \rangle = \sigma_x^2 (1 + \xi_D) + 
 &\frac{\sigma_x^2 \xi_D}{[1+(2 \Lambda t)^2]^3}
 \exp \left[ \frac{-(2 \Lambda t)^2 \xi_D}{1 + (2 \Lambda t)^2} \right]
 \\ &\times \left( \cos(\Phi_2)
 - 6 \Lambda t \sin(\Phi_2)
 - 12 (\Lambda t)^2 \cos(\Phi_2)
 + 8 (\Lambda t)^3 \sin(\Phi_2) \right),
 \label{eq:mom2}
 \end{aligned}
\end{equation}
where $\Phi_2 = 2 \omega t + 2 \phi_D + 2 \Lambda t \xi_D / (1 + (2 \Lambda t)^2)$.
This has a similar form to $\langle x \rangle$, but it oscillates about twice
as rapidly, and decays faster as well, as seen in Figure~\ref{fig:x2}.
In fact, the second moment is less
illustrative of the dynamics than the size of the cloud as a function of time
because the position and size of the cloud are more directly measurable.
The standard deviation of the cloud's position distribution, hereafter 
referred to as the ``cloud size'' is $\sigma(t) = \sqrt{\langle x^2 \rangle -
\langle x \rangle^2}$ and an example is shown in Figure~\ref{fig:sig}.  The decay
rates are important indicators of the harmonicity of the trap, and in
an experimental realization, adjusting the trap parameters to minimize
the decay rate leads toward a maximally harmonic trap.

There is an alternative formulation for the first two moments that works
for $ \Lambda t \ll 1 $ and converts the oscillatory part into a single
cosine.  For the first moment, the approximation yields
\begin{equation}
 \langle x \rangle = \frac{\sqrt{2 \xi_D} \sigma_x}{(1+(\Lambda t)^2)^2}
 \mathrm{exp} \left[ \frac{-(\Lambda t)^2 \xi_D}{1+(\Lambda t)^2} \right]
 \cos \left( \omega t + \phi_D + 2 \Lambda t + \frac{\Lambda t \xi_D}
 {1+(\Lambda t)^2} \right).
 \label{eq:mom1approx}
\end{equation}

Of particular note is the argument of the cosine.  The usual $\omega t + \phi_D$ 
is present, and there is a term proportional to $\xi_D$, representing the 
frequency dependence on the strength of the initial sloshing, but the $2 \Lambda 
t$ term is \textit{independent} of the kick strength.  In other words, the 
frequency of oscillation for a cloud is different from that of a single 
particle, with a frequency difference that tends to a nonzero value even as the 
oscillation amplitude approaches zero.  Contrast this with 
Equation~\ref{eq:1part}, where the value of $\Omega$ for a single particle 
approaches $\omega$ as the amplitude tends to zero.

Figure~\ref{fig:x} shows an example of this solution for a set of parameters 
chosen to showcase the behavior of the cloud as it approaches quasi-thermal 
equilibrium.  The center-of-mass oscillations gradually decay, and eventually no 
center-of-mass oscillations will be observed.  As discussed previously, this 
model does not take collisions into account, so the apparent thermalization is 
purely a result of the relative dephasing of parts of the cloud with different 
energies.

Next, the same approximation applied to $\Upsilon_{22}$ leads to an
approximate expression for the second moment of position.
\begin{equation}
\begin{aligned}
 \langle x^2 \rangle = 
 \sigma_x^2(1+\xi_D)+
 &\frac{\sigma_x^2 \xi_D}{(1+(2 \Lambda t)^2)^3}
 \mathrm{exp} \left[ \frac{-(2 \Lambda t)^2 \xi_D}{1+(2 \Lambda t)^2} \right]
 \\ &\times \cos \left[ 2 \omega t + 2 \phi_D + 6 \Lambda t +
 \frac{2 \Lambda t \xi_D}{1+(2 \Lambda t)^2} \right].
 \label{eq:mom2approx}
\end{aligned}
\end{equation}
These two moments can similarly be combined to approximate the cloud size
as a function of time.

\section{Time of Flight} \label{sec:num}

In an experimental setting, it is likely that the atomic clouds will be imaged
after a period of free-fall, in order to increase the size of the cloud and its
observed oscillation amplitude.  Limited analytical understanding of the behavior
of the clouds after time-of-flight is possible, but numerical calculations can 
more easily illuminate these effects.

The effect on the first moment is a change in amplitude and a phase shift 
that depends on the time of flight.  Looking at the form of the first moment, 
Equation~(\ref{eq:mom1}), its general behavior is that of a sinusoid with a 
slowly varying amplitude.  This functional form is approximated by
\begin{equation}
 \langle x \rangle \approx M(t) \cos(\Omega t + \phi_1),
\end{equation}
where $M(t)$ is the decay function, $\Omega$ represents the real oscillation
frequency of the cloud, and $\phi_1$ represents some initial phase.

Taking the derivative of 
this expression, assuming that the variation of $M(t)$ is slow
compared to the sinusoid, results in
\begin{equation}
 \frac{\partial \langle x \rangle}{\partial t} \approx -\Omega M(t) \sin(\Omega t 
+ \phi_1),
\end{equation}
which has the same frequency of oscillation and a similar decay rate, insofar as
the variation of $M(t)$ is slow, to the dynamics without time of flight.
Assuming the $x$ axis is perpendicular to gravity, the horizontal velocity of the
cloud is constant after the trap is released.  The result is an expression
that oscillates at the same frequency as in the trap, but with a different
phase and amplitude that depend on the time of release and the elapsed time of
flight.

The behavior of the cloud size requires more careful consideration.  Since the 
cloud is oscillating in a potential much larger than the cloud, the cloud may
experience dispersion which can cause the cloud to expand more or less 
quickly after the trap is released, depending on the phase of the oscillation 
at the time of release.  This complication makes writing an analytical 
solution, even a highly approximate one, to $\sigma$ after free-fall much 
more difficult.

\begin{figure}
 \centering
 \includegraphics[width=8.6cm]{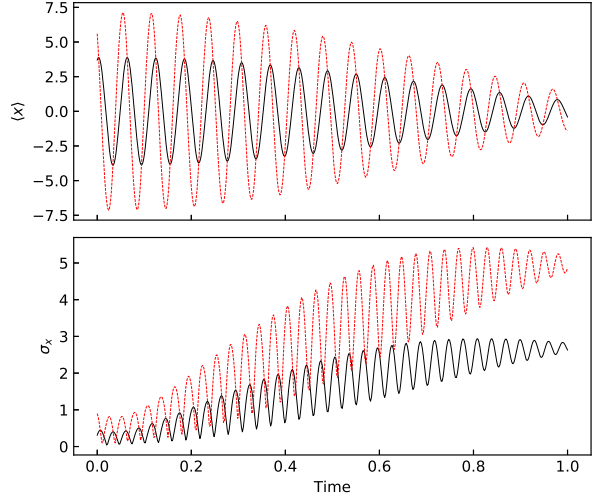}
 \caption{Pictured here are the results of a simple Monte-Carlo numerical 
simulation of sloshing in a harmonic/quartic trap, with trapping parameters 
similar to those used in earlier figures.  Fourth-order Runge-Kutta integration 
was used to evolve the system.  The black curves represent the position and size 
of the cloud without taking time of flight into account.  The dotted curves (red online)
the position and size of the cloud after a short time of flight (0.015).  The 
time of flight exaggerates the oscillations in $\sigma_{x}$, changing the 
shape of the curve.}
 \label{fig:TOFnum}
\end{figure}

Figure \ref{fig:TOFnum} shows the results of a numerical model which shows the 
behavior of $\langle x \rangle$ and $\sigma$, in conditions chosen to be 
similar to the parameters used in Figures (\ref{fig:x}), (\ref{fig:x2}), and 
(\ref{fig:sig}).  We observe that the behavior with and without time of flight
are similar but the time of flight data has a larger amplitude of oscillation,
a larger cloud size in general, and a slightly different size profile as time
goes on, with more exaggerated size oscillations in the middle times.

\section{Conclusions and Outlook} \label{sec:conc}

The oscillations along the weak axis of a small cloud in a perfectly harmonic 
cigar-shaped trap are not expected to decay significantly over time, based on a 
straightforward extension of the calculations in \citet{GO99}.  The addition of 
non-harmonic terms causes the cloud to gradually spread out and reach 
quasi-thermal equilibrium even without interactions, as atoms with different 
energies gradually dephase.  

Our analytical model describes the dynamics of a cloud of neutral atoms in a 
harmonic trap with a quartic perturbation.  We assume that the displacement
of the cloud is less than the magnitude of the quartic parameter $x_4$, that 
mean-field effects are negligible, that collisions are isotropic, 
elastic, and energy-independent, and that the quartic contribution does not 
significantly affect the magnitude of the collisional integral.

With a few minor approximations, the system is analytically solvable. 
The analytical expressions can be used as fitting 
functions for the observed behavior of ensembles of cold atoms in a nearly- 
harmonic trap. The motion of the ensemble's center of mass and the evolution of 
its size permit easy determination of trap anharmonicity, and given the right 
trap architecture, the anharmonic parts can be tuned to some desired value.  
Time-of-flight imaging makes small changes to the observed values of the 
variables but does not qualitatively change the behavior of the ensemble.  Thus, 
even with time-of-flight imaging, the fit parameters for the trap will still 
maintain the same relative trends.

The construction of a trap geometry which can control various polynomial terms 
while minimizing unwanted terms is discussed in \citet{Stickney14}.  The 
specific calculations for a trap allowing adjustment of harmonic and quartic 
terms, while canceling polynomial terms out to sixth order are also detailed 
there.  Dynamic control of the shape of the trap is expected to produce 
additional interesting results.  For instance, because the equilibrium is not a 
true thermal equilibrium, it should be possible to dephase and rephase the cloud 
as it oscillates as long as the collision rate is not too high, and effective 
``pauses'' in the dephasing of the cloud might be made possible by turning off 
the quartic contributions.  These qualities make the dynamically controlled 
trap an interesting topic worthy of experimental examination.

\section{Acknowledgments}

This work was funded by the Air Force Research Laboratory.

\bibliography{for_arxiv.bib}

\end{document}